\documentclass[preprint,showpacs,showkeys,amsmath,amssymb,aps,doublespace]{revtex4}
\usepackage[dvips]{graphicx}
\usepackage{setspace}
\usepackage{amsmath}
\usepackage{amsfonts}
\usepackage{amssymb,color}

\begin{document}

\title{Critical behavior of hard-core lattice gases: Wang-Landau sampling with adaptive windows}

\author{A. G. Cunha-Netto\footnote{e-mail address: agcnetto@fisica.ufmg.br} and R. Dickman\footnote{e-mail address: dickman@fisica.ufmg.br}}
\affiliation{Departamento de F\'{\i}sica, Instituto de Ci\^{e}ncias Exatas,
Universidade Federal de Minas Gerais, C.P.702, 30123-970 Belo Horizonte, Minas Gerais, Brazil}

\begin{abstract}
Critical properties of lattice gases with nearest-neighbor exclusion
are investigated via the adaptive-window Wang-Landau algorithm on the square
and simple cubic lattices, for which the model is known to exhibit an Ising-like
phase transition.
We study the particle density, order parameter, compressibility, Binder cumulant and susceptibility.
Our results show that it is possible to estimate critical exponents using
Wang-Landau sampling with adaptive windows. Finite-size-scaling analysis leads to results
in fair agreement with exact values (in two dimensions) and numerical estimates (in three dimensions).
\end{abstract}



\maketitle

\section{Introduction}

Recently efficient methods for estimating the number of configurations
of classical statistical models have been developed. If the number $\breve{\Omega} (E)$ of
configurations with energy $E$
is determined to sufficient accuracy, many thermodynamic
quantities can be obtained with little further effort, for any desired temperature.
Wang-Landau sampling (WLS) promises to be a simple and reliable approach for
estimating $\breve{\Omega} (E)$ using Monte Carlo simulations \cite{landau,landaupre}.
Critical exponents, the transition temperature (or chemical potential) and other
quantities, such a cumulants, may then be estimated via finite size scaling (FSS)
analysis \cite{fisher,barber,plichske, binney}.
In this paper we test the Wang-Landau algorithm with adaptive windows \cite{adaptive_window},
applying it to the lattice gas with nearest-neighbor exclusion.

Lattice gases have been used extensively as models of simple fluids, and along with the Ising
model have received much attention in equilibrium statistical physics as a prototype for phase transitions.
A particularly simple case is the lattice gas with nearest-neighbor exclusion (NNE), corresponding to an
interparticle potential that is infinite for distances $ \leq 1$ (in units of the lattice constant)
and zero otherwise.  In the absence of an energy scale,
temperature is not a relevant parameter, and the system is termed athermal.  It is known that
on bipartite lattices, the lattice gas with NNE suffers a continuous phase transition between
a disordered phase and an ordered one at a critical value of the density or of the
reduced chemical potential $\mu\equiv\beta\widehat{\mu}$ \cite{gaunt,runnels,blote96,blote02}.
($\widehat{\mu}$ denotes the chemical potential.)
In the ordered phase the occupation fractions
of the two sublattices are unequal.
The grand partition function is

\begin{equation}
\Xi(z,L)=\sum_{N=0}^{N_{max}}z^N\breve{\Omega}(N,L),
\end{equation}
where $z=e^{\mu}$ is the fugacity, $N_{max}$ is the maximum
possible number of particles, and $\breve{\Omega}(N,L)$ the number of distinct configurations
with $N$ particles satisfying the NNE condition, under periodic boundaries.
(On a hypercubic lattice of $L^d$ sites in $d$ dimensions, $N_{max} = L^d/2$ for
$L$ even.)
The order parameter is
the difference between the occupations of sublattices A and B:
\begin{equation}
\phi=\frac{1}{N_{max}}\left\langle \left\vert \sum_{\bf x \in {\rm A}}\sigma_{\bf x}-
\sum_{\bf x \in {\rm B}}\sigma_{\bf x} \right\vert \right\rangle,
\label{eq:ord_par}
\end{equation}
where $\sigma_{\bf x}$ is the indicator variable for occupation of site {\bf x}.

The NNE lattice gas has been studied on various structures: the square \cite{gaunt,blote02,runnels},
triangular \cite{zhang}, simple cubic \cite{blote96}, hexagonal \cite{honeycomb},
body-centered cubic \cite{bodycenter}, and face-centered cubic lattices \cite{facecenter}, and in higher
dimensions \cite{luijten}.  Repulsive lattice gases with exclusion extending to second or further
neighbors have also been studied \cite{runnels,heitor}. Various techniques have been applied to study
the phase transition, including exact enumeration, series expansion (high- and low-density
expansions), the cluster-variation method, transfer matrix analysis, and Monte Carlo simulation.
The model exhibits Ising-like universality on the square and honeycomb lattices,
while on the triangular lattice (Baxter's hard-hexagon model \cite{baxter1,baxter2})
it belongs to the three-state Potts model universality class.

In this paper we calculate the critical
properties of the NNE lattice gas on the square and simple cubic lattices using
the adaptive-window Wang-Landau (AWWL) algorithm, which has been shown to improve
the performance of WLS \cite{adaptive_window,polymer_cpc}.
The critical density and chemical potential, as well critical exponents and the reduced
fourth-order cumulant, are estimated using
FSS analysis.  The balance of this paper is organized as follows.  In Sec. II, the adaptive-window
Wang-Landau algorithm is briefly reviewed. Sec. III contains our results for
the number of configurations (exact enumeration and simulation results),
thermodynamic quantities and critical exponents. A summary is provided in Sec. IV.

\section{Method}

Consider a statistical model with a discrete configuration space, and let $\vartheta$ denote a variable
(or set of variables) characterizing each configuration, such as energy or particle number.  For a given
system size, knowledge of the number $\breve{\Omega}(\vartheta)$ of configurations
(called the ``density of states") for all allowed
values of $\vartheta$ permits one to evaluate the partition function and associated thermal averages
for arbitrary values of the temperature.
Wang-Landau sampling (WLS) \cite{landau,landaupre} furnishes estimates of the configuration
numbers, which we denote by $\Omega(\vartheta)$, reserving $\breve{\Omega}(\vartheta)$ to denote
the {\it exact} values, which are in general unknown.
This is done by performing a random walk in configuration space, with an acceptance probability
proportional to $1/\Omega(\vartheta')$, where $\vartheta'$ denotes the values associated with the
newly generated (or {\it trial}) configuration.
In WLS one aims for equal numbers of visits to each set of allowed values of $\vartheta$, as
reflected in the {\it histogram}, $H(\vartheta)$.

Various strategies have been proposed to improve WLS and optimize its
convergence \cite{zhou_bhatt,adaptive_window,belardinelli1,belardinelli2,zhou}.
In this work we apply one such scheme, adaptive-windows WLS (AWWLS). This method
estimates the density of states by determining the range over which the
histogram has attained the desired degree of uniformity at various stages
of the simulation.

The AWWLS procedure furnishes estimates, $\Omega(N;L)$, of the number of $N$-particle configurations
on a lattice of $L^d$ sites, to within an overall multiplicative factor which is
independent of $N$.  For each accepted $N$-particle
configuration, we update the histogram: $H(N) \to H(N) + 1$.
Since the density of states is not known \textit{a priori}, we set $\Omega(N) = 1$ for all $N$,
at the beginning of each sampling level.
In the simulation, if $N$ and $N+\Delta N$ are the particle numbers
in the current and trial configurations, respectively,
(in practice, $\Delta N=\pm 1$), then the acceptance probability is
\begin{equation}
p(N\rightarrow N+\Delta N)= \min\left[ \frac{\Omega(N)}{\Omega(N+ \Delta N)},1 \right].
\end{equation}
(To simplify the notation we suppress the dependence of $\Omega$ on system size $L$.)

Whenever a move to a configuration with $N$
particles is accepted, the density of states $\Omega(N)$ is updated, multiplying it by a modification
factor $f>1$, so: $\Omega(N) \rightarrow f\cdot\Omega(N)$. If the trial configuration is rejected we update
$\Omega(N)$ (as well as the histogram) of the current $N$ value. As is usual in WLS, the modification
factor is initially set to $f_{0}=e=2.71828...$. After $m=10^{4}$ Monte
Carlo steps we check if the histogram satisfies the flatness criterion on the {\it minimal window}, of width
$W=(N_{max}-N_{min})/n$, beginning with $N_{min}$ \cite{note1}.
The histogram is said to be flat if, for all
levels in the window of interest, $H(N) >0.8 \overline{H}$, where the overline denotes an average
over levels within the proposed window. If it is not flat, we perform an
additional $m$ Monte Carlo steps and check again, repeating until the histogram is flat on
the minimal window. Once this condition is satisfied, we check whether the histogram is
flat on a {\it larger} interval. Thus we define
one window and repeat the procedure on the rest of the range of $N$ values, forming windows for each
stage of sampling.
The window positions depend on the portion
of the histogram that is flat; we include an overlap
of three levels between adjacent windows. This process is repeated until all values of $N$ have been
included in a window with a flat histogram. Then a new stage is initiated: the modification factor is reduced,
$f \rightarrow \sqrt{f}$, and we reset $H(N)=0$ for all $N$. This process is iterated and the
simulation halted when $f -1$ is approximately $10^{-7}$.
As explained in \cite{adaptive_window}, window boundaries are not allowed
to take the same positions
on subsequent stages, to avoid distortions in $\Omega$ that arise when using fixed windows.

For simplicity, two kinds of trial moves are employed: insertion and removal of particles.  Including
particle-displacement moves - for which $N$ does not change - leads to the same results to within
uncertainty. We use
 the R1279 shift register random number generator \cite{landau_binder_book}.

\section{Results}

We study the hard core lattice gas defined above using
AWWLS. The estimates $\Omega(N)$ are used
to calculate $\langle N \rangle$ and var$(N)$ directly; other thermal averages are given by

\begin{equation}
\left\langle A\right\rangle_{\mu}=\frac{\sum_{N}\left\langle A\right\rangle_{N}\Omega(N)e^{\mu N}}{\Xi}.
\label{eq:can_av}
\end{equation}
Here $\left\langle A\right\rangle_{N}$ is the {\it microcanonical} average of quantity $A$
over all configurations having exactly $N$ particles, which must also be estimated
during the simulation.
The development of reliable methods for estimating microcanonical averages is an important open
problem \cite{pmoliveira}.  In the WL procedure, all configurations having the same $N$ should
occur with the same probability, so that, in principle, the microcanonical average
$\left\langle A\right\rangle_{N}$ should be taken over all accepted configurations
having exactly $N$ particles, with equal weights.  We nevertheless obtain better results
if we restrict the microcanonical averages to the later stages of the sampling.
Specifically, the averages $\left\langle \phi\right\rangle_N$, $\left\langle \phi^2\right\rangle_N$
and $\left\langle \phi^4\right\rangle_N$
calculated using all $N$-particle configurations accepted during the simulation yield estimates
for critical exponents that deviate significantly from their expected Ising model values.
Such deviations are not observed
for systems with $L\lesssim 100$, but do appear for larger sizes.
Similar problems were found in studies of spin models
using WLS \cite{malakis2005}.
The results for microcanonical averages improve when we restrict the sample
to configurations accepted in the later stages of the simulation, i.e., for $f\lesssim 1+10^{-4}$.

\subsection{Transfer-matrix analysis}

As a preliminary test of our method, we compare our simulation estimates, $\Omega (N)$, with the results of
an exact enumeration of $\breve{\Omega}(N)$, on a lattice of $8 \times 8$ sites.  The
latter are obtained via a transfer matrix approach.  One begins by enumerating the allowed
configurations $\{c_1,...,c_{\cal M}\}$ on a ring of $L$ sites, and storing the number $n(c_j)$
of particles in each configuration.  An ${\cal M} \times {\cal M}$ matrix ${\cal T}$
is then constructed, with ${\cal T}(c_j,c_j) = 1$ if adjacent rings may assume configurations $c_i$
and $c_j$ without violating the NNE condition, and ${\cal T}(c_i,c_j) = 0$ if the condition is
violated.  Then the allowed configurations on an $L \times L$ lattice with periodic boundaries
are those sequences $\{c_1, c_2,..., c_L\}$ of ring configurations satisfying
${\cal T} (c_1,c_2) {\cal T} (c_2,c_3) \cdots {\cal T}(c_{L-1},c_L) {\cal T}(c_L,c_1) = 1$.
The resulting numbers of configurations for $L=8$ are listed in Table \ref{tab:g}.

To compare our simulation estimates against the exact enumeration, the former must be normalized,
as simulation in fact provides $\Gamma(N) \equiv \alpha \Omega(N)$, with $\alpha$ an unknown constant,
independent of $N$.  To eliminate $\alpha$ we multiply the simulation estimates by a factor $\lambda$,
varying $\lambda$ so as to minimize $\sum_N [\lambda \Gamma (N) - \breve{\Omega}(N)]^2$.
This procedure is applied to each of the fifteen independent simulation studies, leading to
the estimates and uncertainties listed in the Table.
The relative error in estimating $\ln \breve{\Omega}$ is very small except near
the minimum and maximum occupations.  Even in the worst case,
$N=N_{max} = 32$, the relative error in $\ln{\breve{\Omega}}$ is $\approx 0.6\%$.
If sampling errors were restricted to
these regimes for larger system sizes, the effect on estimates for critical properties would be negligible.

\begin{table}[thb!]
\singlespacing
\centering
\begin{tabular}{lllll}
\hline
$N$ ~~ & ~~ $\breve{\Omega}(N)$      ~~ & ~~ $\ln\breve{\Omega}(N)$ ~~ & ~~ $\ln\Omega(N)$  ~~ & ~~ $ 10^3\varepsilon(N)$ \\ \hline
$ 0$ ~~ & ~~ $1$    ~~ & ~~ $            0.0        $ ~~ & ~~ $ 0.010(14)   $ ~~ & ~~ $  -    $ \\
$ 1$ ~~ & ~~ $64$    ~~ & ~~ $           4.15888    $ ~~ & ~~ $ 4.163(11)   $ ~~ & ~~ $ 0.9989$ \\
$ 2$ ~~ & ~~ $1888$    ~~ & ~~ $         7.54327    $ ~~ & ~~ $ 7.543(10)   $ ~~ & ~~ $-0.0481$ \\
$ 3$ ~~ & ~~ $34112$    ~~ & ~~ $        10.43740   $ ~~ & ~~ $ 10.436(11)  $ ~~ & ~~ $-0.1701$ \\
$ 4$ ~~ & ~~ $423152$    ~~ & ~~ $       12.95549   $ ~~ & ~~ $ 12.954(11)  $ ~~ & ~~ $-0.1295$ \\
$ 5$ ~~ & ~~ $3830016$    ~~ & ~~ $      15.15838   $ ~~ & ~~ $ 15.158(9)   $ ~~ & ~~ $-0.0301$ \\
$ 6$ ~~ & ~~ $26249184$    ~~ & ~~ $     17.08315   $ ~~ & ~~ $ 17.086(9)   $ ~~ & ~~ $ 0.1501$ \\
$ 7$ ~~ & ~~ $139580160$    ~~ & ~~ $    18.75415   $ ~~ & ~~ $ 18.754(8)   $ ~~ & ~~ $-0.0003$ \\
$ 8$ ~~ & ~~ $585632520$    ~~ & ~~ $    20.18820   $ ~~ & ~~ $ 20.187(7)   $ ~~ & ~~ $-0.0515$ \\
$ 9$ ~~ & ~~ $1962132800$    ~~ & ~~ $   21.39730   $ ~~ & ~~ $ 21.396(7)   $ ~~ & ~~ $-0.0643$ \\
$10$ ~~ & ~~ $5296005568$    ~~ & ~~ $   22.39022   $ ~~ & ~~ $ 22.392(8)   $ ~~ & ~~ $ 0.0576$ \\
$11$ ~~ & ~~ $11591943552$    ~~ & ~~ $  23.17358   $ ~~ & ~~ $ 23.175(7)   $ ~~ & ~~ $ 0.0447$ \\
$12$ ~~ & ~~ $20681906352$    ~~ & ~~ $  23.75253   $ ~~ & ~~ $ 23.755(8)   $ ~~ & ~~ $ 0.1189$ \\
$13$ ~~ & ~~ $30206108416$    ~~ & ~~ $  24.13131   $ ~~ & ~~ $ 24.135(8)   $ ~~ & ~~ $ 0.1510$ \\
$14$ ~~ & ~~ $36251041536$    ~~ & ~~ $  24.31373   $ ~~ & ~~ $ 24.316(8)   $ ~~ & ~~ $ 0.0881$ \\
$15$ ~~ & ~~ $35886874048$    ~~ & ~~ $  24.30364   $ ~~ & ~~ $ 24.306(7)   $ ~~ & ~~ $ 0.0815$ \\
$16$ ~~ & ~~ $29436488660$    ~~ & ~~ $  24.10550   $ ~~ & ~~ $ 24.109(6)   $ ~~ & ~~ $ 0.1273$ \\
$17$ ~~ & ~~ $20127048512$    ~~ & ~~ $  23.72533   $ ~~ & ~~ $ 23.728(6)   $ ~~ & ~~ $ 0.1330$ \\
$18$ ~~ & ~~ $11573937440$    ~~ & ~~ $  23.17202   $ ~~ & ~~ $ 23.174(7)   $ ~~ & ~~ $ 0.0997$ \\
$19$ ~~ & ~~ $5674532608$    ~~ & ~~ $   22.45925   $ ~~ & ~~ $ 22.462(6)   $ ~~ & ~~ $ 0.1247$ \\
$20$ ~~ & ~~ $2420605568$    ~~ & ~~ $   21.60728   $ ~~ & ~~ $ 21.612(5)   $ ~~ & ~~ $ 0.2128$ \\
$21$ ~~ & ~~ $922331136$    ~~ & ~~ $    20.64241   $ ~~ & ~~ $ 20.646(6)   $ ~~ & ~~ $ 0.1918$ \\
$22$ ~~ & ~~ $322239232$    ~~ & ~~ $    19.59080   $ ~~ & ~~ $ 19.586(7)   $ ~~ & ~~ $-0.2539$ \\
$23$ ~~ & ~~ $104747904$    ~~ & ~~ $    18.46707   $ ~~ & ~~ $ 18.459(8)   $ ~~ & ~~ $-0.4439$ \\
$24$ ~~ & ~~ $31534744$    ~~ & ~~ $     17.26660   $ ~~ & ~~ $ 17.253(11)  $ ~~ & ~~ $-0.7674$ \\
$25$ ~~ & ~~ $8617024$    ~~ & ~~ $      15.96925   $ ~~ & ~~ $ 15.956(12)  $ ~~ & ~~ $-0.7995$ \\
$26$ ~~ & ~~ $2080576$    ~~ & ~~ $      14.54816   $ ~~ & ~~ $ 14.539(12)  $ ~~ & ~~ $-0.6443$ \\
$27$ ~~ & ~~ $430848$    ~~ & ~~ $       12.97351   $ ~~ & ~~ $ 12.973(12)  $ ~~ & ~~ $-0.0571$ \\
$28$ ~~ & ~~ $73840$    ~~ & ~~ $        11.20966   $ ~~ & ~~ $ 11.215(13)  $ ~~ & ~~ $ 0.4370$ \\
$29$ ~~ & ~~ $9984$    ~~ & ~~ $         9.20874    $ ~~ & ~~ $ 9.214(15)   $ ~~ & ~~ $ 0.5301$ \\
$30$ ~~ & ~~ $992$    ~~ & ~~ $          6.89972    $ ~~ & ~~ $ 6.900(15)   $ ~~ & ~~ $ 0.0983$ \\
$31$ ~~ & ~~ $64$    ~~ & ~~ $           4.15888    $ ~~ & ~~ $ 4.160(17)   $ ~~ & ~~ $ 0.2085$ \\
$32$ ~~ & ~~ $2$   ~~ & ~~ $             0.69315    $ ~~ & ~~ $0.689(19)    $ ~~ & ~~ $ -6.359$ \\ \hline
\end{tabular}
\caption{Comparison with numerical results (L=8) for the density of states.
The relative error is $\varepsilon(N)=(\ln\Omega(N)-\ln\breve{\Omega}(N))/\ln\breve{\Omega}(N)$.}
\label{tab:g}
\end{table}

\subsection{Square lattice}

In this work we study $18$ system sizes in the range $16\leq L\leq 256$
using AWWLS.
Let $N^*(L)$  be the value of $N$ that
maximizes $\Omega(N)$, and let $N_c(L)$
be the value of $N$ that maximizes the probability distribution
$P(N)=\Omega(N)\exp[\mu_c N]$ at the critical point $\mu_c$. Preliminary
studies on lattices with $L \leq 200$ reveal that $N^* \simeq 0.227L^2$,
whereas $N_c \simeq 0.369 L^2$, so that $N_c \gg N^*$.
As a result, configurations with $N < N^*$ make a negligible contribution
to thermal averages in the neighborhood of the critical point.
To economize processor time we therefore restrict our
high-statistics studies (which extend to $L=256$), to $N$ values
between $N^*(L)$ and $N_{max}=L^2/2$.
For $L=8$, a study with sampling restricted to $N \geq N^* = 14$ yielded
results of equal accuracy as those obtained using
unrestricted sampling, when compared against exact enumeration.

The following observation suggests that a further economy of
processor time could be realized in studies of the critical region.
Let $P_c (N,L) = z_c^N \Omega(N,L)$
be the contribution to the grand canonical partition function due to
the set of all $N$-particle configurations at the critical point, and
let $P^*(L) = \max_N [P_c(N,L)] = P(N_c)$.  For $N$ values such that
$P_c(N,L)/P^*(L) < 10^{-50}$, say, the contribution to $\Xi$ and
thermal averages is negligible.  It therefore seems reasonable
to restrict the sampling to the interval $[N_1, N_2]$
of $N$ values such that
$P_c(N,L)/P^*(L) \geq 10^{-50}$. In practice we use
$N_1=[0.315L^2-131]$ and $N_2=[0.423L^2+129$], where the brackets
denote the largest integer.  For $L \leq 300$, the largest size considered,
the resulting interval is small enough to be
studied using WLS without windows. Surprisingly, restricting
the sampling in this manner yields estimates for critical exponents
that deviate significantly from their expected Ising model values
(for example, we find $\gamma/\nu=2.02(5)$ on the square lattice).
We conclude that restricted sampling
distorts the estimates for the numbers of configurations, and adversely
affects microcanonical averages.

\begin{figure}[!htb]
\includegraphics[clip,angle=0,width=0.8\hsize]{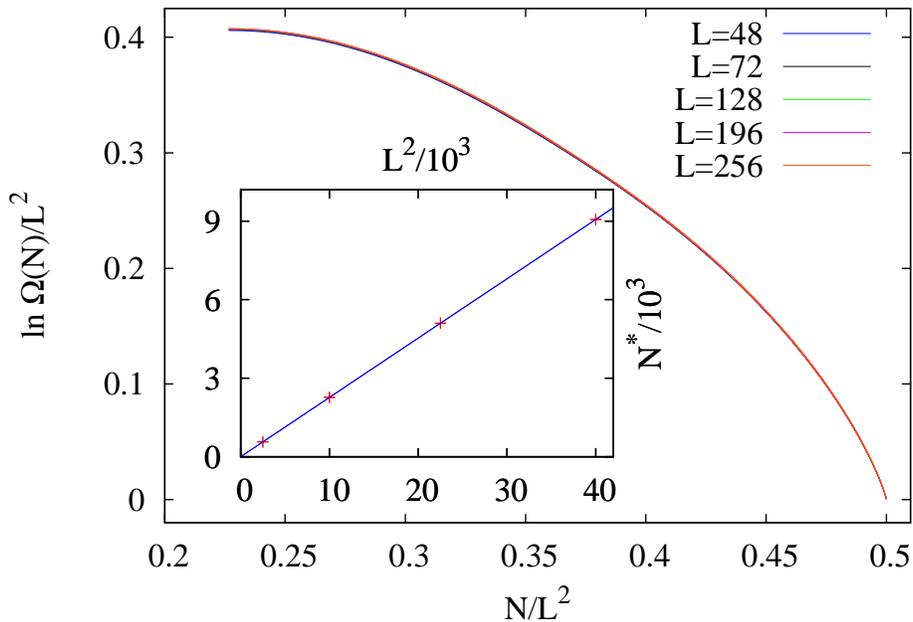}
\caption{(Color online) Square lattice: $\ln \Omega/L^2$ versus
density, system
sizes as indicated. Inset:  $N^*(L)$ versus system size.} \label{fig:coll_lng2d}
\end{figure}

We turn now to the results obtained using the sampling interval
$[N^*(L), N_{max}(L)]$.
Figure \ref{fig:coll_lng2d} shows the
number of configurations versus density $\rho$; the very good data
collapse confirms the expected scaling

\begin{equation}
\Omega (N,L) \simeq \exp \left[ L^d g(\rho) \right]
\label{omega}
\end{equation}

\noindent The inset shows $N^*$ as a
function of system size. Here and below all averages and
uncertainties are obtained using fifteen independent runs.

Two thermodynamic properties used to characterize the transition in
lattice gases are the particle density $\rho (\mu)$ and the
compressibility,
\begin{equation}
\kappa(\mu)=\frac{L^2(\left\langle \rho^2\right\rangle_{\mu}-\left\langle \rho\right\rangle_{\mu}^2)}{\left\langle \rho\right\rangle_{\mu}^2}.
\label{eq:kapa}
\end{equation}

\begin{figure}[!hbt]
\includegraphics[clip,angle=0,width=0.8\hsize]{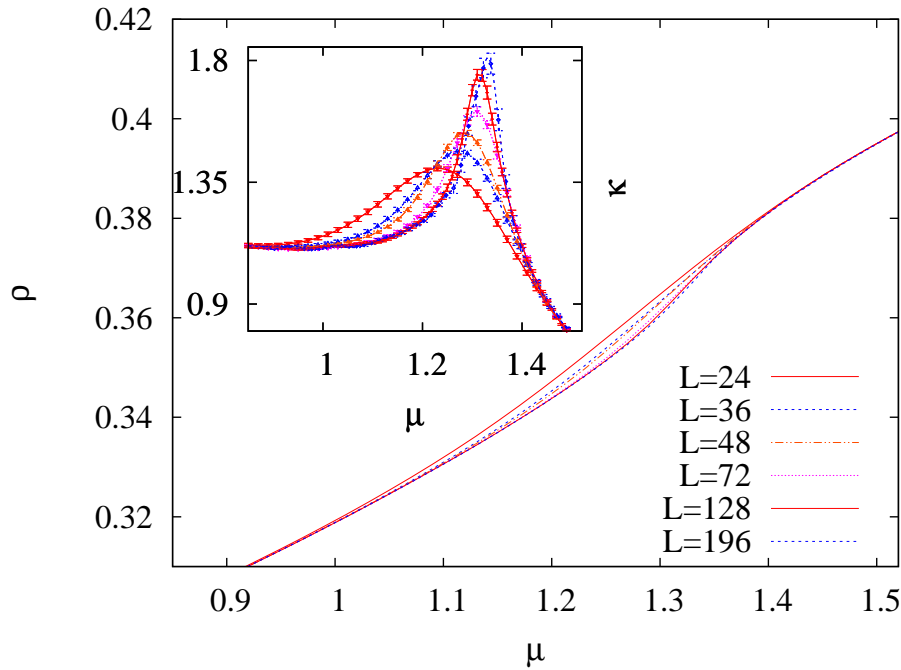}
\caption{(Color online) Square lattice: density versus chemical potential on the
square lattice. Error bars are smaller than the symbols. The
inflection close to the transition point is weak, being
imperceptible for the smaller systems. Inset:
Compressibility versus chemical potential.}
\label{fig:ro_kapa}
\end{figure}

\begin{figure}[!htb]
\includegraphics[clip,angle=0,width=0.8\hsize]{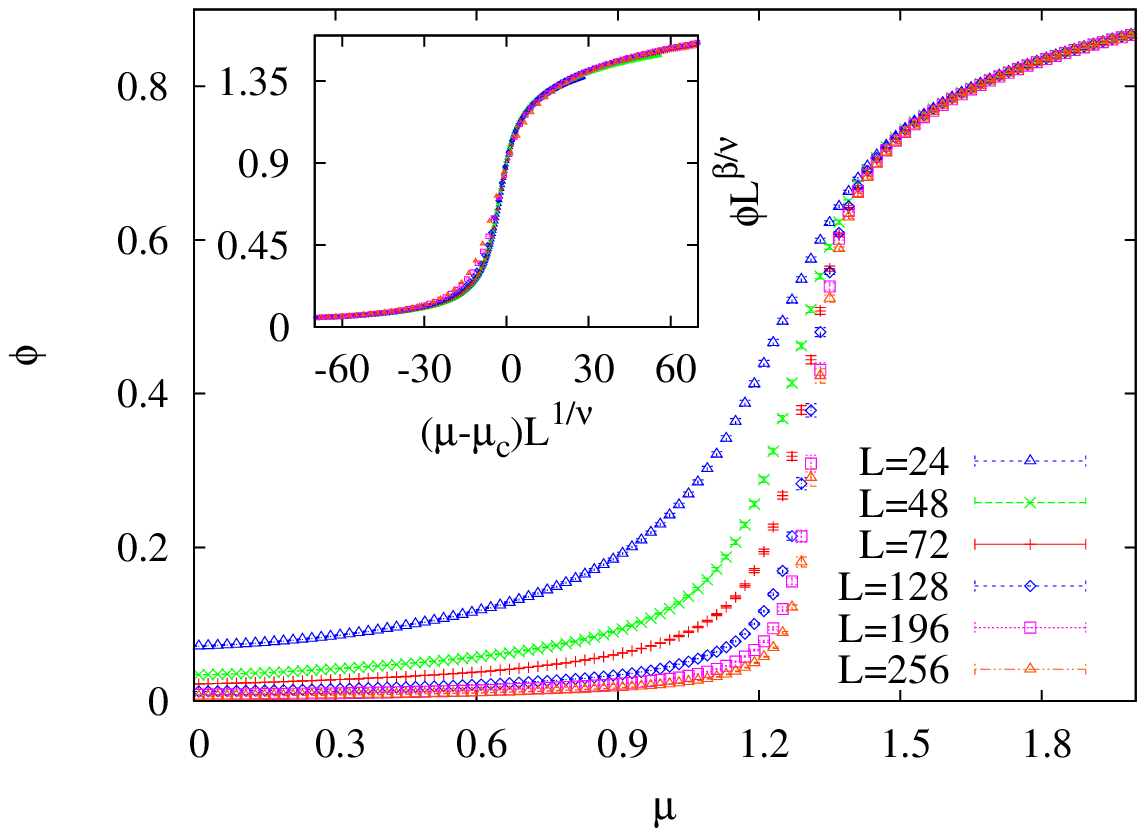}
\includegraphics[clip,angle=0,width=0.8\hsize]{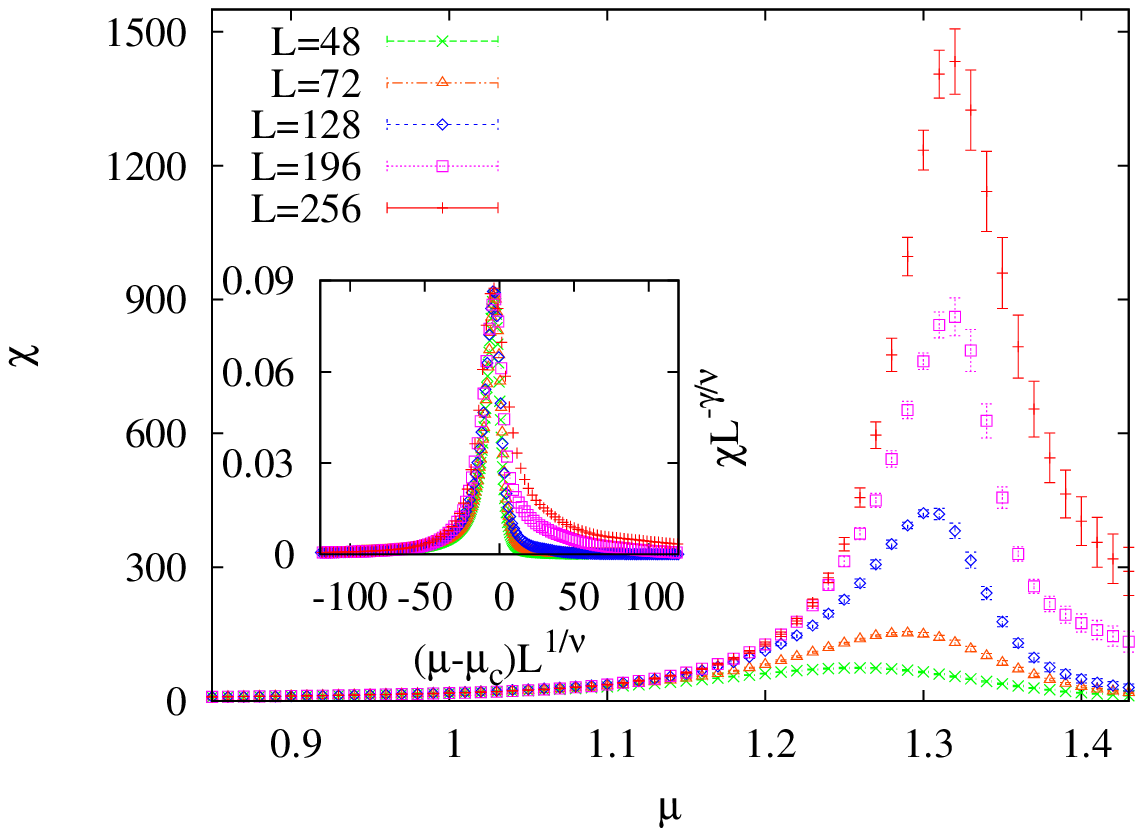}
\caption{(Color online) Square lattice, upper panel: order parameter versus
chemical potential. Lower panel: susceptibility versus
chemical potential. The insets are data-collapse plots.} \label{fig:q_ki}
\end{figure}

\noindent Figure \ref{fig:ro_kapa} shows simulation results for the density $\rho$ and
compressibility $\kappa$ as functions of the chemical potential;
Fig. \ref{fig:q_ki} shows the order parameter, Eq.~(\ref{eq:ord_par}) and the susceptibility,
\begin{equation}
\chi(\mu)=L^2(\left\langle \phi^2\right\rangle_{\mu}-\left\langle \phi\right\rangle_{\mu}^2).
\label{eq:ki}
\end{equation}
The insets in Fig. 3 show the data collapse obtained using the exact critical exponents,
$\gamma/\nu=7/4$ and $\beta/\nu=1/8$, and the high-precision result for
the critical chemical potential obtained by Guo and Bl\"ote \cite{blote02}, $\mu_c=1.33401510027774(1)$.

\begin{figure}[!htb]
\includegraphics[clip,angle=0,width=0.8\hsize]{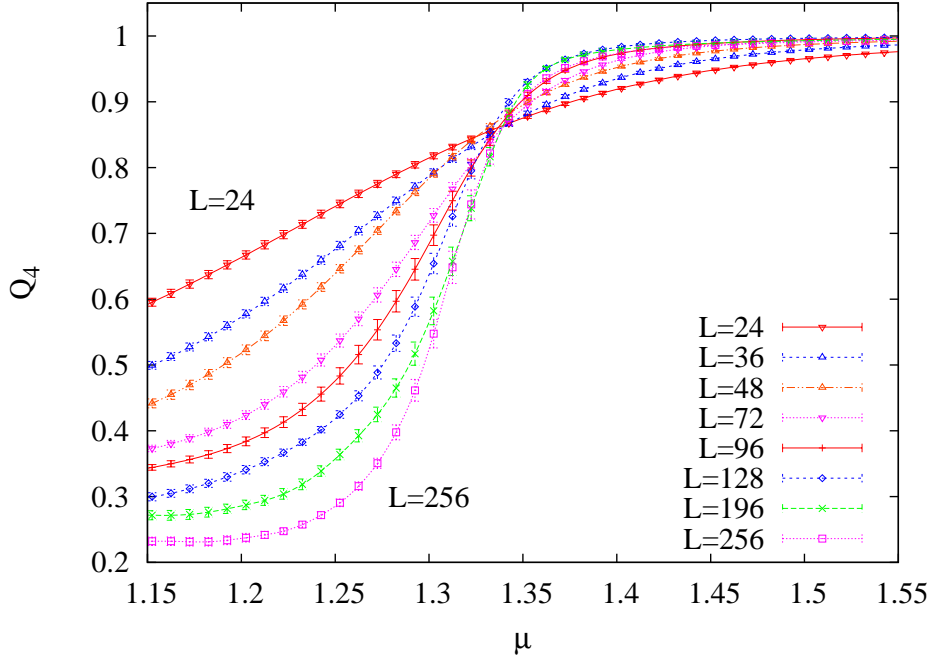}
\caption{(Color online) Square lattice: fourth-order cumulant for system sizes as indicated.}
\label{fig:cum2d}
\end{figure}

We analyzed the dimensionless ratio
$Q_4= \left\langle \phi^2\right\rangle^2/\left\langle \phi^4 \right\rangle$ (Fig.~\ref{fig:cum2d}),
related to Binder's reduced cumulant \cite{binder_cum},
which is expected to take a universal value at
the critical point.
Let $\mu_{c,L_i}$ denote the chemical potential at which $Q_4 (L_i) = Q_4 (L_{i+1}) \equiv Q_{c,L_i}$,
i.e., the crossing between cumulants associated with a pair of successive systems sizes
$L_i$ and $L_{i+1}$, and
let $\overline{L} \equiv\sqrt{L_iL_{i+1}}$
denote the geometric mean of two successive sizes.
It is common to plot
$\mu_{c,L}$ and $Q_{c,L}$ versus $1/\overline{L}^{1/\nu}$ to
estimate the critical chemical potential and cumulant, via extrapolation
to $L \to \infty$.
In the present case, however, we observe
no tendency; all values of $Q_{c,L}$ and $\mu_{c,L}$ agree to within uncertainty.
Averaging over
all values we obtain $\mu_{c}=1.335(3)$ and $Q_{c}=0.852(6)$.
Though of low precision,
these results are consistent with the literature values quoted in Table \ref{tab:results2d}.

\begin{figure}[!hbt]
\includegraphics[clip,angle=0,width=0.8\hsize]{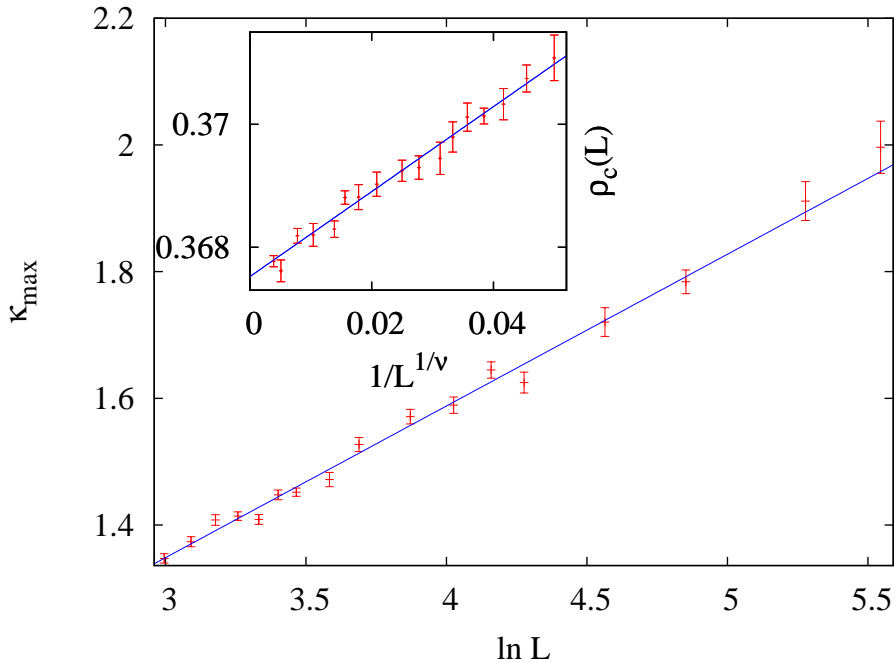}
\caption{(Color online) Square lattice: maximum of the compressibility versus
$\ln L$. The inset shows the density
as a function of $1/L^{1/\nu}$.}
\label{fig:kapamax_roc2d}
\end{figure}

Estimates for the critical chemical potential $\mu_c$ are obtained via analysis of the
chemical potential values associated with the maxima of
the susceptibility and compressibility for each system size.
The extrapolated values, $\mu_c=1.330(1)$ using the susceptibility and
$\mu_c=1.337(2)$ using the compressibility, are obtained using the susceptibility
data for $L=20$ - $196$ and the compressibility data for $L=26$ - $196$.  Pooling our
results, we obtain $\mu_c=1.332(2)$.  Linear extrapolation of the
density $\rho(\mu_c,L)$ versus $1/L$
yields $\rho_c=0.3675(5)$ (see Fig.~\ref{fig:kapamax_roc2d}), inset), consistent with
the critical density reported in \cite{blote02},
$\rho_c=0.3677429990410(3)$. (Using the precise estimate for $\mu_c$ quoted above \cite{blote02},
we obtain $\rho_c=0.36800(5)$.)

Applying FSS analysis to the results for susceptibility for $L=22$ - $256$
yields $\gamma/\nu=1.764(7)$
(see the inset of Fig.~\ref{fig:lnkimax2d}). To estimate $\beta/\nu$ we
analyze $\phi_c(L)$ using the above cited value of $\mu_c$ \cite{blote02};
our data for $L \leq 256$ yield $\beta/\nu=0.123(2)$;
(see Fig.~\ref{fig:lnfilnL2d} inset).
(Using our own less accurate estimate, $\mu_c =1.332(2)$,
we obtain $\beta/\nu=0.130(9)$.) Figure~\ref{fig:kapamax_roc2d} shows
the maximum of the compressibility versus system size; the results are consistent with
$\kappa_{m,L} \sim \ln L$, as expected for a model in the 2d-Ising universality class.
Table \ref{tab:results2d} summarizes our main results.
It is interesting to note that we obtain essentially the same results, $\gamma/\nu=1.764(8)$,
and $\beta/\nu=0.122(3)$, if we exclude the data for the two largest system sizes from the analysis.

\begin{table}[thb!]
\singlespacing
\centering
\begin{tabular}{lll}
\hline
      ~~~~~~ & ~~ Present work ~~ & ~~  Literature values  \\ \hline
$\mu_c$ ~~ & ~~ $1.332(2)$ ~~ & ~~ $1.33401510027774(1)$\footnotemark[1] \\
$Q_c$ ~~ & ~~ $0.852(6)$   ~~ & ~~ $0.856$\footnotemark[2]; $0.855(1)$\footnotemark[3]; $0.85625(5)$\footnotemark[4]  \\
$\rho_{c}$ ~~ & ~~ $0.3675(5)$ ~~ & ~~ $0.3677429990410(3)$\footnotemark[1] \\
$\gamma/\nu$ ~~ & ~~ $1.762(8)$    ~~ & ~~ $7/4$ (exact) \\
$\beta/\nu$ ~~ & ~~ $0.123(2)$    ~~ & ~~ $1/8$ (exact) \\
\hline
\end{tabular}
\footnotetext[1]{Guo and Bl\"ote\cite{blote02}}
\footnotetext[2]{Burkhardt and Derrida\cite{burkhardt}}
\footnotetext[3]{Nicolaides and Bruce\cite{nicolaides}}
\footnotetext[4]{Kamieniarz and Bl\"ote\cite{kamieniarz}}
\caption{Critical values for the square lattice
obtained via WLS with adaptive windows.
The results from \cite{nicolaides} were obtained using Monte Carlo simulations
while Refs. \cite{blote02,burkhardt,kamieniarz}
use a transfer-matrix technique.}
\label{tab:results2d}
\end{table}

\subsection{Cubic lattice}

\begin{figure}[!hbt]
\includegraphics[clip,angle=0,width=0.8\hsize]{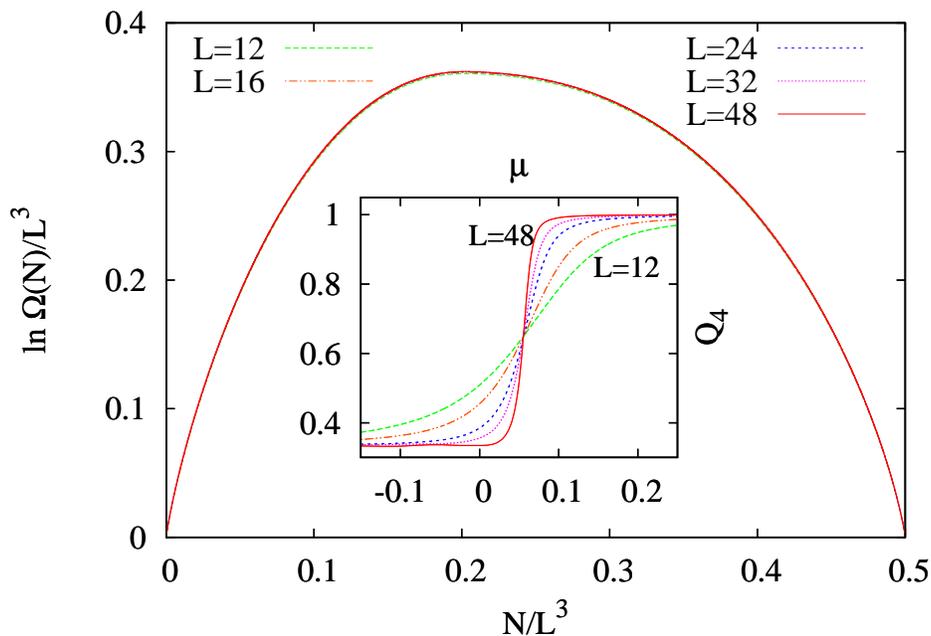}
\caption{(Color online) Simple cubic lattice: $\Omega(N)$ versus density,
system sizes as indicated. Inset: Fourth-order cumulant.}
\label{fig:lng_cum3d}
\end{figure}

\begin{figure}[!hbt]
\includegraphics[clip,angle=0,width=0.8\hsize]{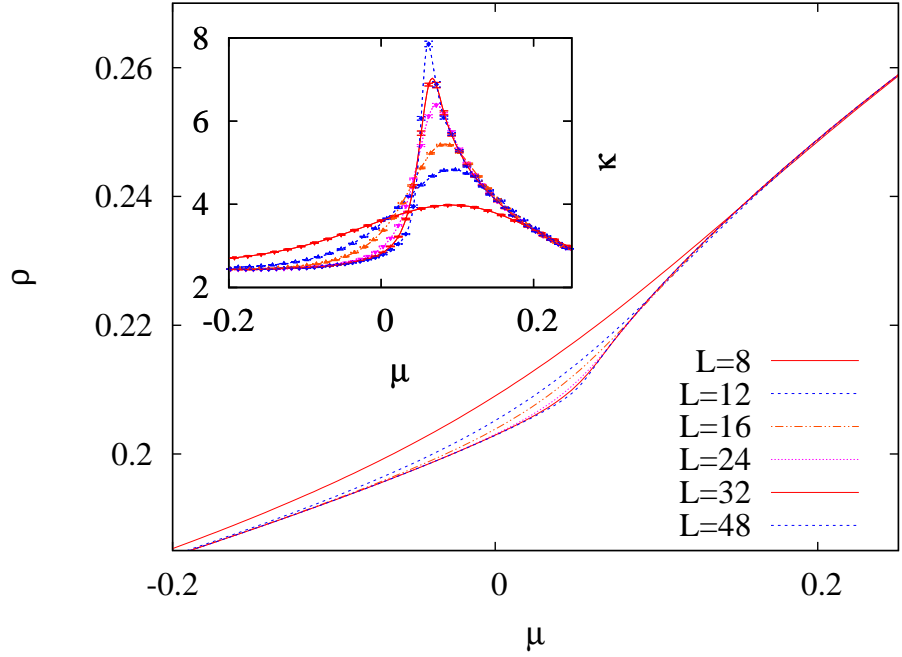}
\caption{(Color online) Simple cubic lattice: density versus chemical potential.
Inset: Compressibility versus chemical potential.}
\label{fig:ro_kapa3d}
\end{figure}

\begin{figure}[!hbt]
\includegraphics[clip,angle=0,width=0.8\hsize]{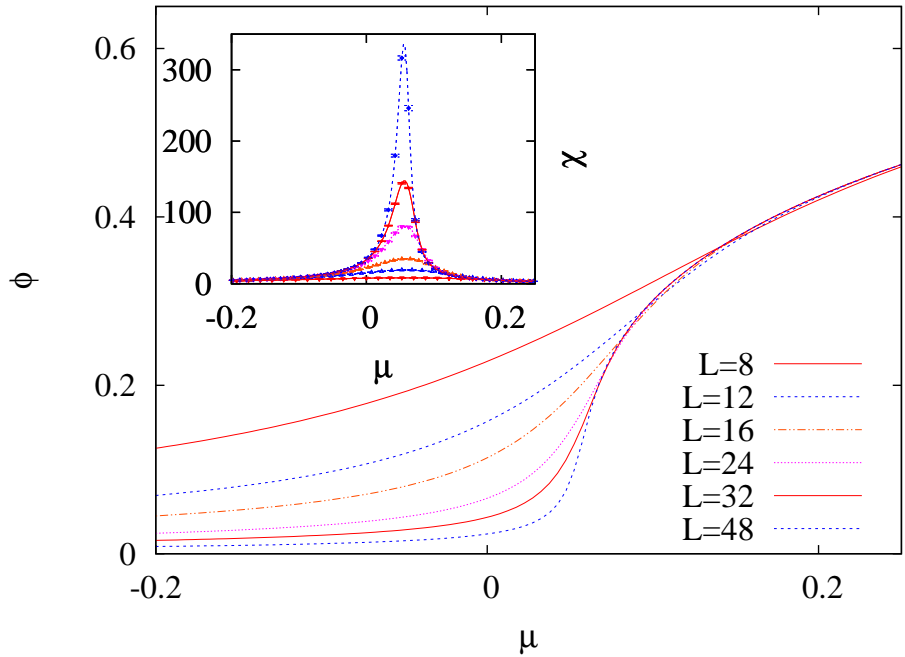}
\caption{(Color online) Simple cubic lattice: Order parameter versus chemical potential.
Inset: Susceptibility versus chemical potential.}
\label{fig:q_ki3d}
\end{figure}

We apply AWWLS to the NNE lattice gas on the
simple cubic lattice, in system sizes
$L=8$, 10, 12, 14, 16, 18, 20, 22, 24,
28, 32, 40 and 48.  In this case we sample the
{\it full range} of $N$ values, using adaptive windows
as described above; for $L=48$, we use approximately fifty windows.
Figure \ref{fig:lng_cum3d} shows $\Omega(N)$,
again verifying Eq.~(\ref{omega}).
The density and compressibility
are plotted versus chemical potential in Fig.~\ref{fig:ro_kapa3d},
while Fig. \ref{fig:q_ki3d} shows the order parameter
and susceptibility, and the inset of
Fig.~\ref{fig:lng_cum3d} the fourth-order cumulant.

Using the critical exponent $\nu=0.6301(4)$ \cite{pelissetto} we plot the values
$\mu_c(L)$ (corresponding to the maxima of the susceptibility and the compressibility) versus
$1/L^{1/\nu}$. Extrapolation of the data for $L > 14$
yields $\mu_c=0.05516(9)$ using the susceptibility, while the compressibility data (for $L > 24$)
yield $\mu_c=0.0567(2)$.
(It is not surprising that the critical value obtained using the compressibility
is less precise than that found using the susceptibility, as the former
exhibits a weaker singularity than the latter.)
Using our best estimate $\mu_c=0.05516(9)$ we calculate
$\rho(\mu_c,L)$; linear extrapolation (for $L > 24$)
versus $1/L^{1/\nu}$ yields $\rho_c=0.21082(5)$
(If we instead use the estimate $\mu_c=0.05443(7)$ \cite{blote96}, we find
$\rho_c=0.21058(5)$).

Proceeding as in the case of the square lattice,
we estimate the critical chemical potential $\mu_{c}$ and the critical
moment ratio $Q_c$. Plotting $\mu_{c,L}$ against $1/\overline{L}^{1/\nu}$ we obtain
$\mu_{c}=0.0552(7)$ via linear extrapolation. This is consistent with previous results
\cite{blote96} which found $\mu_c=0.05443(7)$.
Extrapolation of $Q_{c,L}$ as a function of $1/\overline{L}^{1/\nu}$
yields $Q_{c}=0.652(5)$, somewhat higher than the literature value
(see Table \ref{tab:results3d}).  If we use $\mu_c=0.05443(7)$ \cite{blote96} to calculate $Q_{L}(\mu_c)$,
we observe no significant dependence on $L$; averaging over all values for $L>18$
yields $Q_{c}=0.636(3)$.

\begin{figure}[!hbt]
\includegraphics[clip,angle=0,width=0.8\hsize]{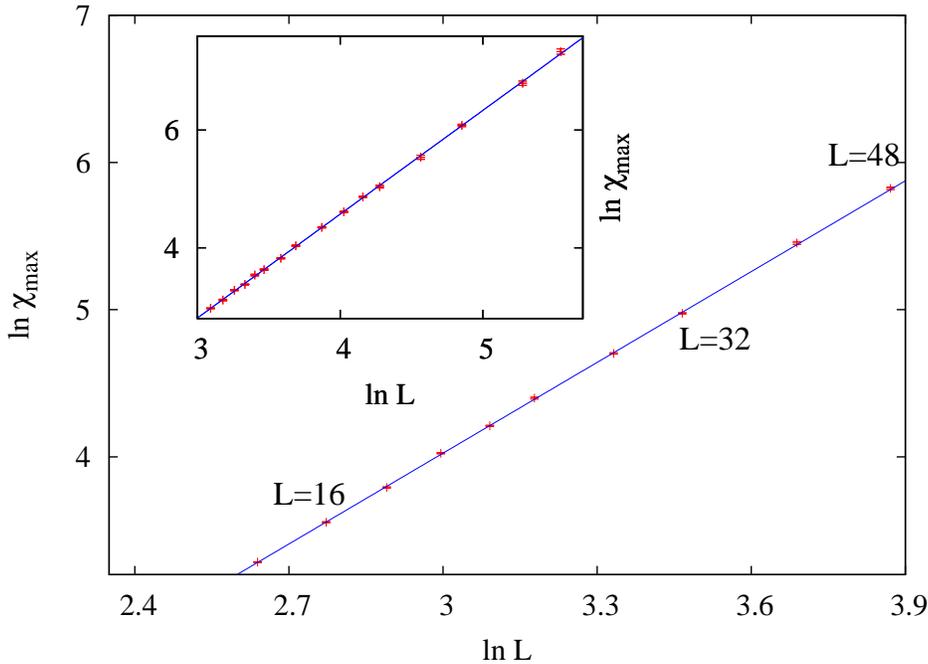}
\caption{(Color online) Maximum of susceptibility versus system size on the square
(inset) and simple cubic
lattices. The solid lines are linear fits used to
estimate $\gamma/\nu$.}
\label{fig:lnkimax2d}
\end{figure}

\begin{figure}[!hbt]
\includegraphics[clip,angle=0,width=0.8\hsize]{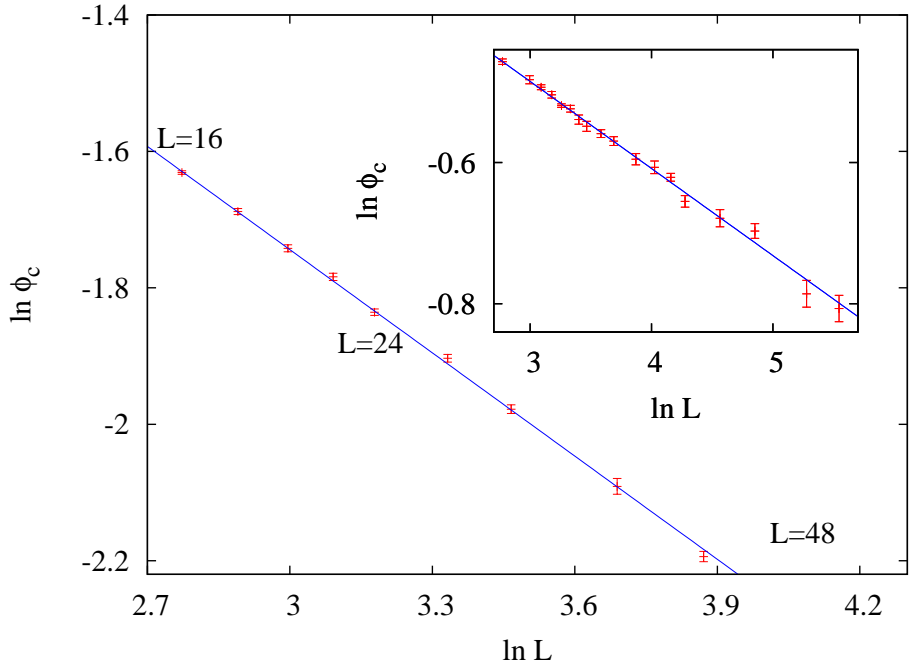}
\caption{(Color online) Critical order parameter versus system size on square (inset)
and simple cubic lattice. The solid lines are linear fits used to obtain $\beta/\nu$.}
\label{fig:lnfilnL2d}
\end{figure}

FSS analysis of the susceptibility furnishes $\gamma/\nu=2.056(6)$ (Fig. \ref{fig:lnkimax2d}).
Using $\mu_c=0.05443(7)$ \cite{blote96}, FSS analysis of the order parameter at $\mu_c$ yields
$\beta/\nu=0.504(8)$ (Fig. \ref{fig:lnfilnL2d}).
(Using our own best estimate, $\mu_c=0.05516(9)$,
we find $\beta/\nu=0.477(7)$). Table \ref{tab:results3d} summarizes our principal results for
the simple cubic lattice. As in the case of the square lattice, our results do not change
significantly if we exclude the data for the two largest system sizes from the analysis.

\begin{table}[thb!]
\singlespacing
\centering
\begin{tabular}{lllll}
\hline
  ~~~ & ~~ Present work  ~~        & ~~ &  Literature values & \\ \hline
$\mu_c$    ~~ & ~~ $0.05516(9)$ ~~ & ~~ $0.05443(7)$\footnotemark[1] ~~ & ~~           $-$                           ~~ & ~~         $-$ \\
$Q_c$      ~~ & ~~ $0.636(3)$   ~~ & ~~ $0.626(4)$\footnotemark[1]   ~~ & ~~           $-$                           ~~ & ~~      $0.6233(4)$\footnotemark[3]  \\
$\rho_{c}$ ~~ & ~~ $0.21082(5)$ ~~ & ~~    $-$                        ~~ & ~~           $-$                          ~~ & ~~         $-$ \\
$\gamma/\nu$ ~~ & ~~ $2.056(6)$ ~~ & ~~ $1.94(2)$\footnotemark[1]    ~~ & ~~        $2.005(6)$ \footnotemark[2]      ~~ & ~~      $1.963(3)$\footnotemark[3] \\
$\beta/\nu$ ~~ & ~~ $0.504(8)$  ~~ & ~~ $0.53(1)$\footnotemark[1]    ~~ & ~~        $0.5002(6)$ \footnotemark[2]     ~~ & ~~     $0.519(2)$\footnotemark[3] \\
$\alpha/\nu$ ~~ & ~~ $0.25(1)$  ~~ & ~~ $0.20(4)$\footnotemark[1]    ~~ & ~~        $-$     ~~ & ~~     $0.174(4)$\footnotemark[3] \\
\hline
\end{tabular}
\footnotetext[1]{Heringa and Bl\"ote \cite{blote96}}
\footnotetext[2]{Garc\'ia and Gonzalo \cite{garcia}}
\footnotetext[3]{Bl\"ote \textit{et al} \cite{blote95}}
\caption{Critical values for the simple cubic lattice
obtained via adaptive-window WLS.
The results from \cite{blote96} were obtained using a cluster algorithm of the NNE lattice gas.
Refs. \cite{garcia} and \cite{blote95} are from high resolution cluster simulations
of the Ising model.}
\label{tab:results3d}
\end{table}

\subsection{Critical behavior of $g(\rho)$}

\begin{figure}[!hbt]
\includegraphics[clip,angle=0,width=0.8\hsize]{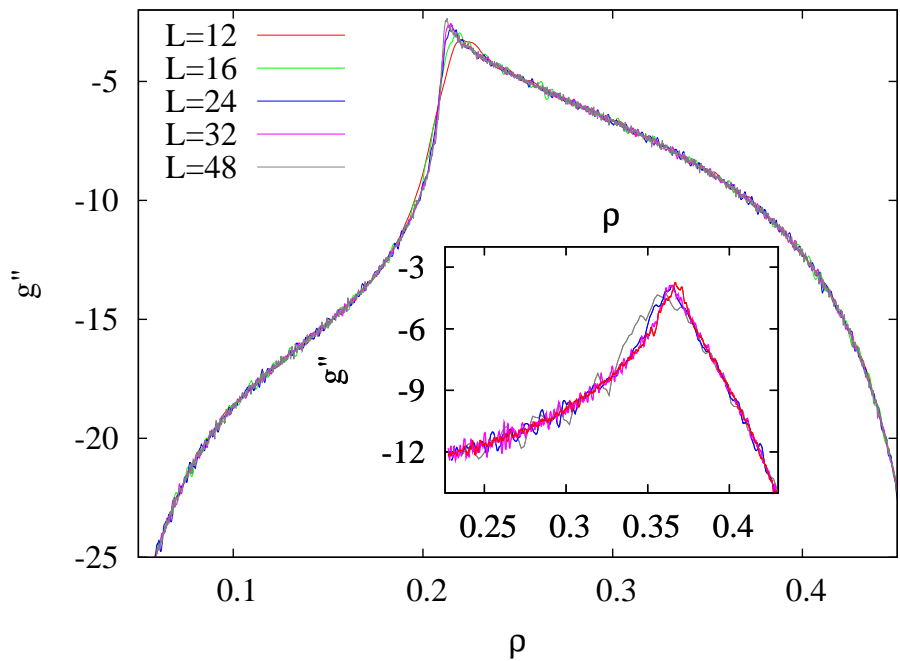}
\caption{Second derivative of $g(\rho)$ on the simple cubic lattice,
system sizes $L=12$, 16, 24, 32 and 48.
The inset is a similar plot for the square lattice for system sizes $L=24$, 48, 128 and 196.}
\label{g248}
\end{figure}

As is known \cite{blote96}, the compressibility of the NNE lattice gas diverges as
$\kappa \sim |\tilde{\mu}|^{-\alpha}$ in the vicinity of the critical point.  (Here $\tilde{\mu}
= (\mu - \mu_c)/\mu_c$ is the reduced chemical potential.)  On the other hand it is easy to
show that $\kappa \propto 1/|g''(\rho)|$, where $g''$ denotes the second derivative of $g$
(defined in Eq. (\ref{omega}) with respect
to $\rho$.  Thus the singularity in the compressibility is reflected in a singularity in $g$.
While plots of $g$ versus $\rho$ appear quite smooth (see Figs. \ref{fig:coll_lng2d} and \ref{fig:lng_cum3d}),
the second derivative does indeed exhibit a singularity near $\rho_c$.  To obtain $g''$ we perform
quadratic fits to $g(N)$ on windows of $b$ successive $N$ values.  We choose $b$ large enough to
eliminate small-scale fluctuations, but small enough that the singular
behavior remains evident \cite{numbers}.
Despite the rounding incurred by such averaging (in addition, of course, to finite-size
rounding), the data shown in
Fig. \ref{g248} provide a clear indication of a developing singularity.
The figure also shows that the minimum of $|g''|$ appears to approach zero as $L \to \infty$.
Our data, however, are not sufficiently precise
to verify the expected scaling, $|g_{min}''| \sim L^{-\alpha/\nu}$.

\section{Conclusions}

We perform adaptive-window Wang-Landau simulations of the
lattice gas with nearest-neighbor exclusion
on the square and simple cubic lattices. On the square lattice, comparison with an
exact enumeration of configurations for $L=8$
yields excellent agreement. Using finite-size scaling analysis
of data for systems of up $256^2$ sites on the square lattice and $48^3$ sites
on the cubic lattice, we estimate the critical
exponent ratios $\gamma/\nu$ and $\beta/\nu$, and the critical values of
the chemical potential, the density, and the fourth cumulant.
In general, fair agreement is observed with literature values.
In three dimensions, the critical point is obtained with an error of about 1.3\%
compared with previous studies, and exponent ratios and $Q_c$ with an error of about 3.5\%.
The precision of our results is considerably less than that obtained using transfer-matrix
or high-resolution Monte Carlo simulations.
Although this is somewhat disappointing, we note that despite the widespread
interest in Wang-Landau sampling, few studies have been published in which critical
exponents are obtained via this technique \cite{malakis2004,malakis2005,malakis2008,malakis2009,malakis2010}.
Our results confirm that it is possible to obtain reasonably accurate values for critical
exponents and related quantities using Wang-Landau sampling with
adaptive windows.  It thus appears worthwhile to seek further improvements
in the method, in efforts to develop a simple and versatile approach for
studying phase transitions via Monte Carlo simulation.

In this regard two observations seem pertinent.  The first is that restricting sampling
to a subset of densities may worsen the results, even though densities outside this
subset make a negligible contribution to thermal averages.  It appears that
the imposition of reflecting barriers on the random walk in configuration space
somehow distorts the sampling.  The second point is that the quality of the results
furnished by the WLS procedure appears to decay with increasing system size, even while
maintaining the same flatness criterion and schedule of updates of the factor $f$.
Thus, including larger systems sizes in the analysis may not improve results;
more extensive sampling of small and intermediate system sizes may represent a more
effective allocation of computing resources.  We hope to explore the reasons for, and implications
of these observations in future work.

\noindent \section*{Acknowledgments}

We are grateful to CNPq, and Fapemig (Brazil) for financial support.


\begin{thebibliography}{999}
\bibitem{note1} On the square lattice we use $n=4$ for $L \leq 72$, $n=6$ for $L=96$, $n=8$
for $L$ between 100 and 196, and $n=10$ for $L=256$. On the simple cubic lattice
$n=2$ for $L \leq 12$, $n=4$ for $L=16$, $n=6$ for $L$ between 24 and 32, and $n=8$ for $L=48$.
\bibitem{numbers} For the square lattice we use $b=3$ for $L=24$, $b=7$ for $L=48$, 20 for
$L=128$ and 50 for $L=196$; for the simple cubic lattice, $b=10$ for $L=12$ and 16,
$b=20$ for $L=24$, 40 for $L=32$ and 70 for $L=48$.
\end{thebibliography}
\end{document}